\documentclass[usenatbib,usegraphicx]{mn2e}

\usepackage{amsmath}

\title[Thermal emission from WASP-33b]{Thermal emission from WASP-33b, the hottest known planet\thanks{Based on service observations made with the William Herschel Telescope operated on the island of La Palma by the Isaac Newton Group in the Spanish Observatorio del Roque de los Muchachos of the Instituto de Astrof'sica de Canarias.}}

\author[A. M. S. Smith et al.]{A. M. S. Smith$^{1}$\thanks{E-mail:
amss@astro.keele.ac.uk},
D. R. Anderson$^{1}$,
I. Skillen$^{2}$,
A. Collier Cameron$^{3}$,
and B. Smalley$^{1}$\\
$^{1}$Astrophysics Group, Lennard-Jones Laboratories, Keele University, Keele, Staffordshire ST5 5BG\\
$^{2}$Isaac Newton Group of Telescopes, Apartado de Correos 321, E-38700 Santa Cruz de la Palma, Tenerife, Spain\\ 
$^{3}$SUPA, School of Physics and Astronomy, University of St Andrews, North Haugh, St Andrews, Fife KY16 9SS \\
} 

\begin{document}
\maketitle

\begin{abstract}

We report ground-based observations at 0.91~\micron~of the occultation of the hot Jupiter WASP-33b by its A5 host star. We measure the planet to be $0.109 \pm 0.030$ per cent as bright as its host star at 0.91 \micron. This corresponds to a brightness temperature, $T_{\rm{B}} = 3620^{+ 200}_{- 250}$~K, significantly higher than the zero-albedo equilibrium temperature for both isotropic re-radiation ($2750 \pm 37$~K) and uniform day-side only re-radiation ($3271 \pm 44$~K), but consistent with the zero-redistribution temperature ($3515 \pm 47$~K). This indicates that the heat redistribution from the day-side of WASP-33b to the night side is inefficient, and further suggests that there is immediate re-radiation, and therefore little or no redistribution, of heat within the day-side. We also detected the stellar pulsations of WASP-33, which we model as the sum of four sinusoids, with periods of between 42 and 77 minutes and amplitudes of 0.5 to 1.5 mmag.

\end{abstract}
\section{Introduction}

Extra-solar planets which transit their host stars are also likely to exhibit occultations (secondary eclipses), in which the planet passes out of sight behind the star. Observing an occultation allows measurement of the planet's emergent flux. This flux consists of thermal emission and reflected light, but is strongly dominated by thermal emission in the infrared wavelength regime. The first detections of thermal emission from an exoplanet were made using the {\em Spitzer} space telescope (\citealt{Charbonneau-etal05}; \citealt{Deming-etal05}), since when the occultations of many more planets have been observed at wavelengths of 3.6~\micron~and greater with {\em Spitzer}.

Complementary to {\em Spitzer} observations at 3.6, 4.5, 5.8, 8 and 24 \micron, are ground-based observations made in the near infrared. Such observations are important because they extend the observed planetary spectral energy distribution (SED) to lower wavelengths, and to the peak of the SED, which is around 1~\micron~for the shortest-period, hottest planets.

Measurement of a planet's occultation depth allows the brightness temperature to be determined, which leads to an evaluation of how efficiently heat is redistributed from the day-side to the night-side of the planet. 

A planet's observed SED (from occultation observations at several different wavelengths) allows the chemical composition of the atmosphere to be inferred (e.g. \citealt{swain10}). Detection of a thermal inversion in the atmosphere is also possible; observations to date indicate two classes of hot-Jupiter atmosphere, those with and those without such an inversion. One possible explanation for the existence of these two classes is that provided by \cite{knutson_stellar_activity} who argue that chromospherically active stars host planets with no temperature inversion.

Occultation observations also allow refinement of the orbital parameters of a planetary system, particularly the orbital eccentricity, which is related to both the timing of the occultation and its duration.

WASP-33 (HD 15082) was first identified as a transiting planet candidate by \cite{Christian-etal06}, but was identified as fast rotator, which rules out the use of precision radial velocity measurements to confirm its planetary nature and to help characterise the system. Instead, \cite{wasp33}, hereafter C10, used line-profile tomography during transit to confirm WASP-33b as the first known planet to orbit an A-type star.

Given the nature of its host star and its very short orbital period (1.22~d), WASP-33b has the largest equilibrium temperature ($T_{\mathrm{eql}} \approx  2700$~K, assuming zero-albedo and uniform heat redistribution; C10) of any known exoplanet. WASP-33b is an excellent target for ground-based occultation observations, due to the large predicted planet-to-star flux ratio. Planets with similarly large brightness temperatures include WASP-19b ($T_{\mathrm{eql}} \approx 2000$~K; \citealt{w19}), occultations of which have been observed from the ground in $H$-band and $K$-band (\citealt{w19hawki}; \citealt{w19gibson}) and WASP-12b ($T_{\mathrm{eql}} \approx 2500$~K; \citealt{w12}), observed from the ground in the $ K_S$-, $H$- and $J$-bands by \cite{w12croll}.

C10 also reported that the stellar line profiles show evidence for non-radial pulsations, and suggest that the star may be a $\gamma$~Dor-type variable. Further, photometric, evidence for these pulsations is presented in \cite{herrero}, hereafter H11, who suggest that WASP-33 is a $\delta$~Scuti-type variable.

In this paper, we present ground-based observations of the occultation of WASP-33b at 0.9\micron.

\section{Observations and data reduction}

We observed the occultation of WASP-33b on 2010 October 29/30 using the Auxiliary Camera (ACAM) of the William Herschel Telescope (WHT). The ACAM CCD covers a field-of-view of $8 \arcmin \times 8\arcmin$ with $2048\times 2048$ pixels, giving a plate scale of $0.25 \arcsec$ per pixel. The target was positioned on the CCD such that three relatively bright comparison stars are present in the images. A total of 203 images were taken from 22:43 to 05:55 UT.

The observations were conducted using a narrow-band S[III] filter (ING filter \#S9077), which has a central wavelength of 9077~\AA, and a FWHM of 54~\AA. The telescope was defocussed to spread the light from the target over a large number of pixels, thus reducing noise associated with imperfect flat-fielding and allowing longer integration times. Exposure times of 100, 120 and 150~s were used. The conditions were not photometric; thin cirrus cloud affected the observations for the duration of the night.

Aperture photometry was performed on the flat-field and bias corrected images, using the photometry pipeline of \cite{Southworth-wasp5}, which is based around the {\sc astrolib / aper} routine\footnote{The {\sc astrolib} subroutine library is distributed by NASA. For 
further details see http://idlastro.gsfc.nasa.gov}. The pointing was monitored by cross-correlating each image with a reference image, and the apertures shifted to follow the stars. 

\begin{figure}
\includegraphics[angle=270, width=84mm]{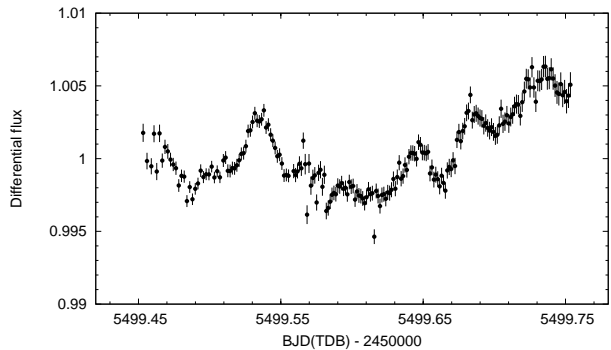}
\caption{Normalised raw light curve.}
\label{fig:rawlc}
\end{figure}

Photometry was performed using a range of different aperture radii, from 8 to 65 pixels. A radius of 50 pixels was chosen to maximise the target signal-to noise, and ensure that a large fraction ($>95$ per cent) of the target flux is included in the aperture. The flux from the three comparison stars was combined to form an ensemble reference star, approximately 0.5 magnitudes fainter than WASP-33. The resulting differential light curve is shown in Fig. \ref{fig:rawlc}.

\section{Data analysis}

In addition to an apparent eclipse-like signal (centred about $\mathrm{JD} \approx 2455499.58$ as expected, assuming a circular orbit), the raw light curve (Fig. \ref{fig:rawlc}) exhibits large systematic effects, specifically (i) a series of sine-like oscillations, which we attribute to the stellar pulsations for which evidence was presented by C10 and H11; and (ii) an increase in flux across the second half of the light curve.

We analyse the new occultation data alongside the SuperWASP, Keele and JGT light curves and the radial velocities of C10, and the Montsec Observatory light curve of H11. This global analysis is performed using the Markov Chain Monte Carlo (MCMC) code described in \cite{Cameron-etal07}, \cite{Enoch10} and \cite{wasp-30}.

In addition to fitting the occultation, we also simultaneously fit functions describing the stellar pulsations and systematic effects present in the raw light curve (Fig. \ref{fig:rawlc}). The de-trending within the MCMC code is done using the {\sc svdfit} routine \citep{numrec}. Initially, we neglected the stellar pulsations, and attempted only to remove the instrumental effects. We tried de-trending with several parameters: airmass, time, sky background amplitude, and the position of the target on the CCD; we also tried de-trending using various combinations of parameters and various functional forms for the fit to each parameter. We used the Bayesian information criterion (BIC) to discriminate between the different models. In determining the best de-trending model we did not re-scale the error bars on any of the data. The lowest BIC value was produced when the data were de-trended with a quadratic function of the sky background,
\begin{equation}
g = a_0 + a_1 f_{\mathrm{sky}} + a_2 f_{\mathrm{sky}}^2
\label{eqn:bg}
\end{equation}
where $f_{\mathrm{sky}}$ is the sky background flux and $a_0$, $a_1$, and $a_2$ are the coefficients we fit for.

The light curve resulting from our de-trending with sky background, while an improvement upon the raw light curve, still exhibits the sine-like variations present in the raw light curve which we attribute to stellar pulsations. It is apparent the signature of these pulsations is multi-periodic; a model consisting of a single sine curve produced a very poor fit to the data. 

\begin{figure}
\includegraphics[angle=0, width=84mm]{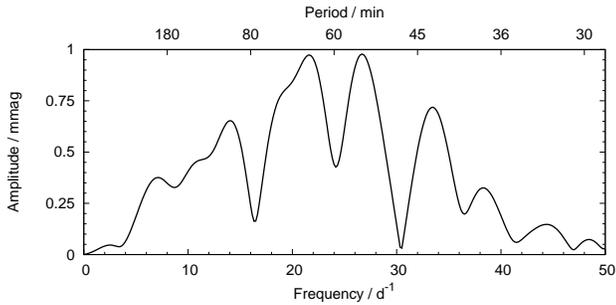}
\caption{Periodogram output from {\sc Period04}, for the out-of-occultation data.}
\label{fig:pgram}
\end{figure}

We conducted a Fourier analysis on the out-of-occultation data using the {\sc Period04} time-series analysis software package \citep{period04}. This reveals three frequencies which are deemed significant (with a signal-to-noise ratio greater than 4). These frequencies correspond to periods of $53.99\pm0.29$ min, $76.90 \pm 0.62$ min and $41.85 \pm 0.31$ min (see periodogram, Fig. \ref{fig:pgram}). The 68-min period of H11 is present in our periodogram, but with a slightly lower amplitude than the 54-min period. When prewhitening with the 54-min period, the the significance of the 68-min period is much reduced, suggesting that one of these two periods may be an alias of the other. When, instead we used the H11 period and amplitude as the starting point for a {\sc Period04} analysis yields a different set of frequencies, but a poorer overall fit to our data. We also tried a period search of our data combined with the existing photometry, but the results from this were inconclusive.

We then fitted for either a single pulsation period (that of H11), three pulsation periods (from our {\sc Period04} analysis) or four pulsations periods (our periods and that of H11) with an MCMC code (see below). Comparing the BIC values of the resulting models we found that the 4-period model ($\Delta BIC = 0$) is preferred to both the 3-period ($\Delta BIC = 391$) and the 1-period ($\Delta BIC = 1057$) models.

The four periods and uncertainties were input to a version of our MCMC code, modified to include the four pulsation periods as proposal parameters. The periods were, therefore, allowed to vary but Gaussian priors were applied to prevent the periods from drifting too far from the fitted periods. The priors were applied by means of a Bayesian penalty added to our merit function ($\chi^2$), given by
\begin{equation}
BP_{P_i} = \frac{(P_i - P_{i,0})^2}{ \sigma^2_{P_{i,0}}}
\end{equation}
where $P_{i,0}$ and $\sigma^2_{P_{i,0}}$ are the initial value and uncertainty of a particular period, $P_i$, respectively. The amplitudes and phases of each of these sine waves were fitted using {\sc svdfit}, along with the coefficients of Equation \ref{eqn:bg}. The function fitted is of the form
\begin{equation}
h = g(f_{\mathrm{sky}}) + \displaystyle\sum\limits_{i=1}^4 b_i \sin\left(\frac{2 \pi t}{P_i}\right) + c_i \cos\left(\frac{2 \pi t}{P_i}\right)
\label{eqn:bg_sine}
\end{equation}
where $g(f_{\mathrm{sky}})$ is Equation \ref{eqn:bg}, $t$ is time, and $b_i$ and $c_i$ are the coefficients we determine in the fit. This is equivalent to
\begin{equation}
h = g(f_{\mathrm{sky}})  + \displaystyle\sum\limits_{i=1}^4 A_i \sin\left(\frac{2 \pi t}{P_i} + \delta_i \right)
\label{eqn:bg_sine2}
\end{equation}
where $A_i$ are the amplitudes, and $\delta_i$ the phase offsets of each sine term.

\begin{table}
\begin{center}
\caption{Comparison of de-trending models, $g$, (see Equation \ref{eqn:bg_sine}). $\Delta BIC$ is the value of $BIC$ relative to that of the model with the lowest value of $BIC$}
\label{tab:bic}
\begin{tabular}{lc}
\hline
Functional form of $g$ &  $\Delta BIC$  \\
\hline
$a_0 + a_1 f_{\mathrm{sky}} + a_2 f_{\mathrm{sky}}^2$                 & 0 \\
$a_0 + a_1 f_{\mathrm{sky}} $                                                             & 4 \\
$a_0 + a_1 \delta t + a_2 (\delta t)^2 $                                               & 14 \\
$a_0 + a_1 f_{\mathrm{sky}}^2$                                                          & 123 \\
$a_0 + a_1 \sec{z} + a_2 \sec^2{z}$                                                   & 118 \\

\hline
\end{tabular}
\end{center}
$f_{\mathrm{sky}}$ is the sky background flux, $z$ is the zenith distance of the target, and $\delta t$ is the time since the start of the observations.
\medskip
\end{table}

\begin{table*} 
\caption{System parameters for WASP-33} 
\label{tab:params} 
\begin{tabular*}{0.9\textwidth}{@{\extracolsep{\fill}}lcc} 
\hline 
Parameter (Unit) & (C10) & This work \\
\hline 
\\
Orbital period, $P$ (d) & $1.2198669 \pm 0.0000012$ &  	  	$1.21986983 \pm 0.00000045$\\
Epoch of mid-transit, $T_{\rm c}$ (HJD) & $2454163.22373 \pm 0.00026 $&  $	2454590.17948 \pm 0.00028$\\
Transit and occultation durations, $T_{\rm 14} = T_{\rm 58}$ (d) & -   &  	$0.11224 \pm 0.00084$\\
Transit and occultation ingress and egress &&\\
~~~~~~~~~durations, $T_{\rm 12} = T_{\rm 34} = T_{\rm 56} = T_{\rm 78} $ (d) & -   &  	0.01149$^{+ 0.00097}_{- 0.00084}$\\
Transit depth, $\Delta F=R_{\rm P}^{2}$/R$_{*}^{2}$ & $0.01136 \pm 0.00019$  &  	0.01041$^{+ 0.00023}_{- 0.00021}$\\
Impact parameter, $b$ & 0.16$^{+ 0.10}_{- 0.12}$ &  	0.32$^{+ 0.09}_{- 0.12}$\\
Orbital inclination, $i$ ($^\circ$) \medskip & $87.7 \pm 1.8$ &  	  	84.9$^{+ 2.1}_{- 1.7}$\\
Orbital eccentricity, $e$ & -  &  	0 (fixed) \\
Occultation depth, $\Delta F_{0.91\,\micron}$ & - &  $	0.00109 \pm 0.00030$\\
Stellar effective temperature, $T_{\rm *, eff}$ (K) & $7430 \pm 100$  &  	$7430 \pm 100$\\
Stellar metallicity, [M/H] & $0.1 \pm 0.2$ & $0.1 \pm 0.2$\\
Stellar mass, $M_{\rm *}$ ($M_{\rm \odot}$) & $1.495 \pm 0.031$  & $ 	1.512 \pm 0.040$\\
Stellar radius, $R_{\rm *}$ ($R_{\rm \odot}$) & $1.444 \pm 0.034$  &  	1.512$^{+ 0.060}_{- 0.054}$\\
Stellar surface gravity, $\log g_{*}$ (cgs) & $ 4.3 \pm 0.2$ &    	4.258$^{+ 0.027}_{- 0.029}$\\
Stellar density, $\rho_{\rm *}$ ($\rho_{\rm \odot}$) \medskip & $ 0.497 \pm 0.024$ &  	$0.437 \pm 0.045$\\
Planet mass, $M_{\rm P}$ ($M_{\rm Jup}$) & $< 4.1$  & $< 4.59$ (3-$\sigma$) \\
Planet radius, $R_{\rm P}$ ($R_{\rm Jup}$) & $ 1.497 \pm 0.045$  &  	1.501$^{+ 0.073}_{- 0.064}$\\
Orbital major semi-axis, $a$ (AU)  & $0.02555 \pm 0.00017$   & $ 	0.02565 \pm 0.00023$\\
Planet equilibrium temperature, $T_{\rm {eql} (A=0, f=1)}$ (K) & - &  	  	$2750 \pm 37$\\
Planet equilibrium temperature, $T_{\rm {eql} (A=0, f=2)}$ (K) & - &  	  	$3271 \pm 44$\\
Planet equilibrium temperature, $T_{\rm {eql} (A=0, f=\frac{8}{3})}$ (K) & - & 	$3515 \pm 47$\\
Planet brightness temperature, $T_{\rm B, 0.91\micron}$ (K) & - & $3620^{+ 200}_{- 250}$\\
Pulsation period 1,  $P_{\rm{P1}}$ (min) & - & 53.5$^{+ 0.9}_{- 0.6}$\\
Pulsation period 2,  $P_{\rm{P2}}$ (min) & - & 76.3$^{+ 2.6}_{- 4.0}$\\
Pulsation period 3,  $P_{\rm{P3}}$ (min) & - & $41.9 \pm 0.5$\\
Pulsation period 4,  $P_{\rm{P4}}$ (min) & - & 66.6$^{+ 2.0}_{- 2.1}$\\
Pulsation amplitude 1,  $A_{\rm{P1}}$ (mmag) & - &  1.479 $ \pm 0.069$\\
Pulsation amplitude 2,  $A_{\rm{P2}}$ (mmag) & - &  0.567 $ \pm 0.134$\\
Pulsation amplitude 3,  $A_{\rm{P3}}$ (mmag) & - &  0.766 $ \pm 0.115$\\
Pulsation amplitude 4,  $A_{\rm{P4}}$ (mmag) & - &  0.605 $ \pm 0.105$\\
\\ 
\hline 
\end{tabular*} 
\end{table*}

We then again tried different forms of Equation \ref{eqn:bg} to de-trend the data, again finding that a quadratic in sky background results in the lowest BIC value (Table \ref{tab:bic}). Once we had selected the functional form of our trend model, we performed a further MCMC analysis, this time scaling the error bars of each photometric dataset so as to obtain a reduced $\chi^2$ of unity.

We initially fitted for the orbital eccentricity and the resulting solution is significantly eccentric ($e = 0.174^{+0.058}_{-0.039}$), but $e\cos\omega$ is close to, and consistent with, zero ($e\cos \omega = 0.0027^{+0.0079}_{-0.0033}$). The argument of periastron, $\omega$, in this solution is $-89.0^{+3.0}_{-1.2} \degr$, i.e. the major axis of the orbit is aligned almost exactly with our line-of-sight. We suggest that this is an improbable configuration, and is caused by a degeneracy in fitting the occultation and the stellar pulsations. $e\cos\omega$ is constrained by the timing of the occultation, whereas the duration of the occultation informs us about $e\sin\omega$.

\begin{figure}
\includegraphics[angle=0, width=84mm]{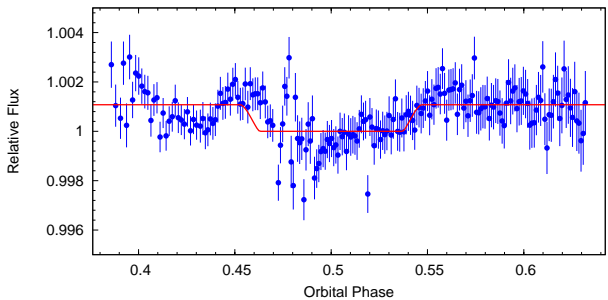}
\caption{Occultation photometry, de-trended for sky background and with stellar pulsations subtracted, normalised to unity in occultation (blue points), with best-fitting model overplotted (red curve).}
\label{fig:occ}
\end{figure}

Furthermore, in the case of WASP-33, the shape of the radial velocity curve places almost no constraint on the eccentricity (because the RVs are imprecise due to the nature of the star). This allows the fitted occultation duration to deviate from the transit duration to a large extent. The light curve, after de-trending for sky background and stellar pulsations still clearly contains correlated (`red') noise, which is likely to be caused by pulsation activity at further periods (Fig. \ref{fig:occ}). It is this red noise that mimics the ingress and egress of the occultation and results in an occultation duration significantly shorter than the transit duration, and an argument of periastron close to $-90\degr$.

The eccentric solution requires that the planet be moving more slowly during transit than for the circular solution, and this in turn requires that the stellar radius is smaller in the eccentric case. This results in a stellar density ($\rho_{\rm *} = 0.75^{+0.16}_{-0.12}~\rho_{\rm \odot}$) which is greater than that of a main-sequence star with the mass of WASP-33 ($\rho_{\rm *} \approx 0.5~\rho_{\rm \odot}$; Gray 2008). The stellar density of our circular solution, however, lies very close to that prediction - $\rho_{\rm *} = 0.44 \pm 0.05~\rho_{\rm \odot}$. \nocite{gray-book}

Although short-period planets have been found in significantly eccentric orbits (e.g. WASP-14b, \citealt{wasp14}; \citealt{husnoo10}), the orbit of WASP-33b is expected to have been circularised by tidal interactions with the star. These are anticipated to be particularly strong because of the relatively large radii of planet and star, and the small orbital separation.

Using Equation (1) of \cite{Jackson08}, we calculate the circularisation time-scale, $\tau_e = \left(\frac{1}{e} \frac{de}{dt}\right)^{-1}$ using the values of $M_{\rm *}$, $R_{\rm *}$, $a$, and $R_{\rm P}$ quoted in Table \ref{tab:params}, and $M_{\rm P} = 4.1M_{\rm Jup}$ (the upper limit established by C10, which will maximise the calculated circularisation time-scale). We obtain 
\begin{equation}
\tau_e = \left(\frac{0.557}{\left (\frac{Q_{\rm P}}{10^{5.5}}\right)} + \frac{0.024}{\left (\frac{Q_{\rm *}}{10^{6.5}}\right)}\right)^{-1}~\mathrm{Myr,}
\end{equation}
where $Q_{\rm P}$ and $Q_{\rm *}$ are the tidal dissipation parameters for the planet and star respectively. Adopting $Q_{\rm P} = 10^{5.5}$ and $Q_{\rm *} = 10^{6.5}$ (the best-fitting values from the study of \citealt{Jackson08}), we find $\tau_e = 1.72$ Myr. This time-scale is an order of magnitude less than the 25 Myr main-sequence age estimate of C10, and a very small fraction of the 500 Myr upper limit to the age. Although this suggests that there has been ample time for the orbit to have been circularised, $\tau_e$ is affected by large uncertainties in the $Q$ parameters, so we cannot say for certain that this is the case.

C10 determined that the orbital axis of WASP-33b is misaligned with respect to the stellar spin axis, such that the planet orbits in a retrograde sense. This suggests that the system may not exist in a dynamically relaxed state, and the orbit may be eccentric. The time-scale for co-planarisation of the orbit, however, is very much longer than that for its circularisation. Using Equation (14) of \cite{Winn05} we obtain the characteristic time-scale, $\tau_\psi$, for significant change in the angle $\psi$ between the planetary orbital angular momentum vector and the stellar rotation angular momentum vector,
\begin{equation}
\tau_\psi = 11.4 \left(\frac{Q_{\rm *}}{10^{6.5}}\right)~\mathrm{Gyr.}
\end{equation}
Here, we have assumed values of 0.01 for the apsidal motion constant, $\sqrt{0.1}$ for the dimensionless radius of gyration, and a planet mass of 4.1 $M_{\rm Jup}$. The time-scale increases with decreasing planet mass. It is therefore perfectly reasonable to suppose that the orbit of WASP-33b is circular, despite the fact that it is retrograde.

Furthermore, the empirical evidence provided by other planetary systems lends support to the circular hypothesis. \cite{Pont_ecc} present an analysis of the eccentricities of a large number of hot Jupiters, determining that the only known planets with significantly detected, large eccentricities are massive and have relatively long-period orbits. Several planets with large spin-orbit misalignments are known to have near-circular orbits, such as WASP-17b which has a retrograde orbit \citep{wasp17}, and an orbital eccentricity close to zero, $e = 0.028^{+ 0.015}_{- 0.018}$ \citep{wasp17_spitzer}.

\begin{figure}
\includegraphics[angle=0, width=84mm]{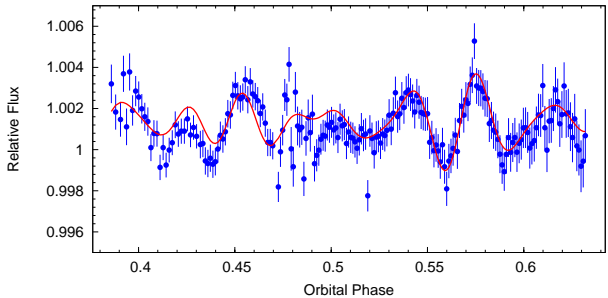}
\caption{Stellar pulsations. Photometry, with sky background and occultation models subtracted (blue points), overplotted with our best-fitting pulsation model (red curve).}
\label{fig:pulse}
\end{figure}

All of the above evidence leads us to rule out the eccentric solution, and to take the  approach (advocated by e.g. \citealt{w44_45_46}) of adopting a circular solution. The best-fitting parameters  for this circular model are presented in Table \ref{tab:params}. The fit to the pulsations is shown in Fig. \ref{fig:pulse} and the fit to the occultation is shown in Fig. \ref{fig:occ}.

\begin{figure}
\includegraphics[angle=0, width=84mm]{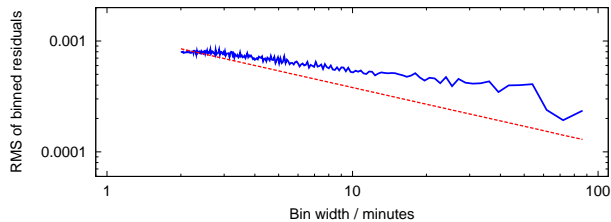}
\caption{RMS of the binned residuals as a function of bin size (solid, blue curve). The dashed, red curve is the expected RMS for noise that is purely uncorrelated, corresponding to the RMS of the unbinned data divided by the square root of the number of points per bin.}
\label{fig:rms}
\end{figure}

We assessed the presence of correlated noise in the occultation light curve residuals, by plotting the RMS of the binned residuals to the best-fitting model as a function of bin width (Fig. \ref{fig:rms}). The RMS of the binned residuals deviates significantly from the uncorrelated noise expectation, indicating significant levels of red noise are present in the light curve. This is likely to be due to the presence of pulsation signals in addition to those we fitted. The results of a Fourier analysis to determine whether any further periodicity is present in the data is inconclusive, in part due to the limited baseline of the data. To determine whether fitting an occultation model to the data is justified, we performed an F-test to compare our best-fitting occultation model with a constant-flux model. The resulting $p=0.013$ (i.e. there is only a $\sim 1 \%$ chance of our occultation fit arising if there were no underlying occultation) indicates that the occultation model is preferred at the $>$ 95 per cent level.

The remaining correlated noise present in the occultation light curve means that the uncertainty on the occultation depth is likely to have been underestimated. To calculate a realistic uncertainty, we employ the residual permutation or `prayer bead' method (e.g. \citealt{Gillon07}). We took the light curve residuals and shifted each residual to the subsequent observation before adding back the trend, pulsation and occultation models. This process was repeated, shifting each residual two observations, and so on, resulting in a total of $n_{\mathrm{obs}} - 1 = 202$ light curves, each of which have the time-structure of the correlated noise preserved. A separate MCMC analysis was performed on each of these light curves.

We then calculated the median of the best-fitting occultation depths and pulsation periods from the new fits, as well as the uncertainty limits which enclose 68 per cent of the values around the median. The uncertainties on these parameters quoted in Table \ref{tab:params} are those calculated from the permutation of the residuals. In all cases, the median values from the residual permutation analysis are consistent with, and close to the best-fitting values from our single MCMC analysis. As expected, the uncertainties on these parameters derived from the permutation analysis are larger. The original occultation depth we found was $0.00108 \pm 0.016$, whereas the value from the residual permutation is $0.00109 \pm 0.030$.

\section{Discussion}

To calculate the brightness temperature of the planet, we modelled the star using a synthetic spectrum of an A5 star \cite{Pickles98} normalised to the integrated flux of a 7430~K (Table \ref{tab:params}) black-body emitter, and the planet as a black-body of temperature $T_{\rm{B}}$. We then defined the measured occultation depth as the product of the ratio of the bandpass-integrated planet and star fluxes and the planet-to-star area ratio (e.g. \citealt{Charbonneau-etal05}). We calculate a brightness temperature, $T_{\rm{B}} = 3620^{+200}_{-250}$~K. The uncertainty on $T_{\rm{B}}$ only accounts for the uncertainty on the measured occultation depth, which is the dominant source of error. Additional, smaller sources of uncertainty on the ratio of planetary and stellar areas, and the stellar effective temperature are not accounted for.

The equilibrium temperature for a zero-albedo planet is given by $T_{\rm P, A=0} =  f^{\frac{1}{4}} ~T_{*,\rm eff}~\sqrt{\frac{R_*}{2a}}$, where $f = 1$ indicates isotropic re-radiation over the whole planet (i.e. the redistribution of heat from the day-side to the night-side is fully efficient). The case where the heat is uniformly distributed on the day-side of the planet, but there is no redistribution of heat to the night-side  corresponds to $f = 2$. A third case, corresponding to $f = \frac{8}{3}$ (equivalent to $\varepsilon = 0$ and $f = \frac{2}{3}$ in the notations of \citealt{CowanAgol} and \citealt{LM_S} respectively), occurs when the incident radiation is immediately re-radiated and so there is no redistribution of heat even within the day-side. Because the hottest region of the day-side is most visible close to occultation, a deeper occultation is observed than if the day-side had a uniform temperature.

The brightness temperature of 3620 K is greater than either the equilibrium temperature for uniform or no redistribution to the night-side (respectively 2750 K and 3271 K, Table \ref{tab:params}). The brightness temperature is consistent with the zero-redistribution ($f=\frac{8}{3}$) case (3515 K), however. This suggests that the heat-transport from the day-side to the night-side is inefficient, and further that the day-side is unlikely to have a uniform temperature. This is in line with observations of other highly irradiated planets, which are also found to have poor heat redistribution efficiencies (e.g. \citealt{CowanAgol}).

\section{Conclusions}

We have detected thermal emission from WASP-33b. We measure the occultation depth at $0.91 \micron$ to be $0.109 \pm 0.030$ per cent, which corresponds to a brightness temperature of $3620^{+200}_{-250}$~K, the hottest such temperature recorded for an exoplanet.

We also detect the non-radial pulsations of the host star in the photometry; and conclude that a multi-periodic solution is required to fit the pulsation signal. Our best-fitting model of the pulsation signal consists of sine terms with periods of 53.5, 76.3, 41.9, and 66.6 min, with amplitudes of 1.5, 0.6, 0.8 and 0.6 mmag respectively. None of these periods corresponds directly to the period of 68 min fitted by H11, but are similar, and our greatest amplitude agrees with their 0.9 mmag. We do however find evidence for the H11 68-min period in our periodogram, and it appears that the 54-min and 68-min periods may be aliases of each other. We do not claim that these periods are definitive, but rather they are the best fit to our data. Our analysis suggests that the pulsations of WASP-33 are complex and multi-periodic in nature. This multi-periodic, $\sim 1$~h, pulsation signature lends support to the conclusion of H11 that WASP-33 is a $\delta$~Scuti-type variable. Our ability to determine the true periodicity of the pulsations is limited by the relatively short baseline of our data.

Although fitting for the orbital eccentricity returns a significantly non-zero value of $e$, we argue that because $e \cos \omega$ is essentially zero, the eccentric solution is improbable. Furthermore, the eccentric solution is incompatible with the stellar analysis of C10, so we therefore adopt a circular model for the orbit.

High signal-to-noise photometry with a longer baseline is required to study better the pulsations, to determine the complete nature of the periodicity and to investigate whether, as H11 suggest, there is evidence for star-planet interactions in the WASP-33 system. Such a characterisation of the pulsation periods would also allow the pulsation signal to be more cleanly subtracted from occultation photometry. This, in turn, would allow a  more precise measurement of the occultation depth and the unresolved questions concerning the orbital eccentricity to be answered. Measurements of the occultation depth at different wavelengths to the one presented here are required in order to construct a spectral energy distribution and hence to characterise the planetary atmosphere.

\section{Acknowledgements}
The authors wish to extend their thanks to E. Herrero for supplying the Montsec Observatory transit light curve, and to the anonymous referee for suggestions that improved the final manuscript.

\bibliographystyle{mn2e_fixed}
\bibliography{iau_journals,refs2}
\bsp

\end{document}